\def\year{2022}\relax
\documentclass[letterpaper]{article} 
\usepackage{aaai22}  
\usepackage{times}  
\usepackage{helvet}  
\usepackage{courier}  
\usepackage[hyphens]{url}  
\usepackage{graphicx} 
\usepackage{natbib}  
\usepackage{caption} 
\DeclareCaptionStyle{ruled}{labelfont=normalfont,labelsep=colon,strut=off} 
\usepackage{algorithm}
\usepackage{algorithmic}

\usepackage{color}

\usepackage{newfloat}
\usepackage{listings}
\floatstyle{ruled}
\newfloat{listing}{tb}{lst}{}
\floatname{listing}{Listing}
\title{A Synthetic Prediction Market for Estimating Confidence in Published Work}

\author {
    Sarah Rajtmajer,\textsuperscript{\rm 1}
    Christopher Griffin,\textsuperscript{\rm 1}
    Jian Wu,\textsuperscript{\rm 2}
    Robert Fraleigh,\textsuperscript{\rm 1}
    Laxmaan Balaji,\textsuperscript{\rm 1}
    Anna Squicciarini,\textsuperscript{\rm 1}
    Anthony Kwasnica,\textsuperscript{\rm 1}
    David Pennock,\textsuperscript{\rm 3}
    Michael McLaughlin,\textsuperscript{\rm 1}
    Timothy Fritton,\textsuperscript{\rm 1}
    Nishanth Nakshatri,\textsuperscript{\rm 1}
    Arjun Menon,\textsuperscript{\rm 1}
    Sai Ajay Modukuri,\textsuperscript{\rm 1}
    Rajal Nivargi,\textsuperscript{\rm 1}
    Xin Wei,\textsuperscript{\rm 2}
    C. Lee Giles\textsuperscript{\rm 1}
}
\affiliations {
    \textsuperscript{\rm 1}The Pennsylvania State University\\
    \textsuperscript{\rm 2}Old Dominion University\\
    \textsuperscript{\rm 3}Rutgers University\\
        \{smr48,cxg286,rdf5090,lpb5347,acs20,amk17,mvm7085,tjf115,nzn5185,amm8987,svm6277,rfn5089,clg20\}@psu.edu\\
             \{j1wu,xwei001\}@odu.edu\\
 david.pennock@rutgers.edu
}

\begin{document}

\maketitle

\begin{abstract}
Explainably estimating confidence in published scholarly work offers opportunity for faster and more robust scientific progress. We develop a synthetic prediction market to assess the credibility of published claims in the social and behavioral sciences literature. 
We demonstrate our system and detail our findings using a collection of known replication projects. We suggest that this work lays the foundation for a research agenda that creatively uses AI for peer review.
\end{abstract}

\section{Introduction}

\noindent Concerns about the replicability, robustness and reproducibility of findings in scientific literature have gained widespread attention over the last decade in the social sciences and beyond, including AI \cite{gundersen2018state,henderson2018deep,hutson2018artificial,haibe2020transparency,pineau2021improving}. This attention has been catalyzed by and has likewise motivated a number of large-scale replication projects \cite{open2015estimating,camerer2016evaluating,camerer2018evaluating,klein2014investigating,klein2018many,cova2021estimating} which have reported successful replication rates anywhere between 36\% and 78\% and have further escalated debate of a crisis of confidence in present-day empirical work \cite{baker20161,gilbert2016comment,fanelli2018opinion}.

Given the challenges and significant resources required to run high-powered replication studies, researchers have sought other approaches to assess confidence in published claims and have looked to creative assembly of expert judgement as one opportunity. Initial evidence has supported the promise of prediction markets in this context \cite{dreber2015using,camerer2016evaluating,camerer2018evaluating,forsell2019predicting,gordon2020replication,gordon2021predicting}. 
However, practical deployment of prediction markets to evaluate scientific findings is also limited. They require the coordinated, sustained effort of collections of human experts. 
They typically rely on availability of some measurement of ground truth. That is, participants trade on well-defined, verifiable outcomes determined after market close (although, see \cite{liu2020surrogate} for recent work proposing a surrogate scoring mechanism). 

Another set of limitations centers around the 
shortcomings of human market participants. 
Researchers 
base their 
assessments on 
the work with which they are familiar, the reputations of journals, and similar. 
Their judgements may be influenced by cognitive biases, e.g.,  anchoring, confirmation bias \cite{fraser2021predicting}, and the compounded effects of these biases in market settings are poorly understood. 


We suggest that markets populated by artificial agents 
provide an opportunity to overcome or mitigate many of these limitations. 
Synthetic prediction markets can be deployed rapidly and at scale. 
Artificial agents can have broad access to the literature and metadata at scales far beyond the capacity 
of an individual researcher. 

The system we demonstrate here is a fully synthetic prediction market wherein algorithmic agents (trader bots) are trained and tested on proxy ground truth pulled from existing replication studies. 
Our work is complementary to recent efforts using machine learning for reproducibility prediction \cite{altmejd2019predicting,yang2020estimating,pawel2020probabilistic,wu2021predicting}. Unlike prior approaches the market scores only a subset of the papers in our test set, but accuracy on that subset is very high. The market-based approach affords explainability by way of the record of trades and corresponding relevant features.



\begin{figure*}[h!]
\centering
\includegraphics[width=0.99\textwidth]{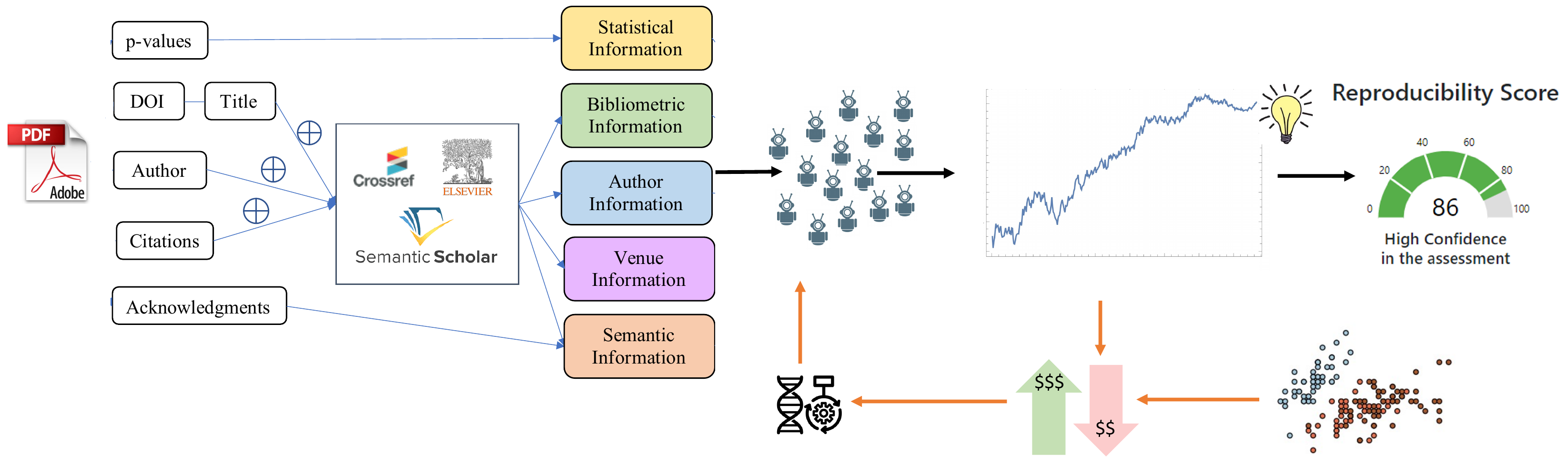} 
\caption{(Black arrows) A scientific paper is processed through the FEXRep feature extraction framework. Features are shared with the agents who purchase assets corresponding to 
binary outcomes of a notional replication study of the primary claim of that paper. The price of 
these assets at market close is an indicator of confidence in the claim. 
(Orange arrows) During training, agents purchase assets corresponding to claims drawn from prior replication projects for which ground truth is known. At the close of each market, some agents profit and others lose money. An evolutionary algorithm is used to update the population.}
\label{fig:schematic}
\end{figure*}

\section{System}

The prototype system is built around two primary modules, namely, a feature extraction pipeline and the synthetic market. Outputs of feature extraction are provided to agents which populate the market during train and test (Figure \ref{fig:schematic}).

\subsection{Feature Extraction Pipeline}

The Feature EXtraction framework for Replicability prediction (FEXRep) extracts five categories of features related to a given scholarly preprint or published paper and its metadata: biblometric, venue-related, author-related, statistical and semantic information. At present, 41 total features are extracted, ranging from p values and sample size to number of authors and acknowledgement of funding. Further detail is provided in \cite{wu2021predicting}.

In the prototype system, all features represent paper-level information. Ongoing efforts are expanding extraction to incorporate features at the claim-level. This will allow for individual assessment of multiple claims within the same paper, rather than the current approach which scores the primary claim of the paper as it is asserted in the abstract.

\subsection{Synthetic Market}

Agents in the market are initialized with a fixed amount of cash and provided with the set of extracted features representing a paper in question. Agents may purchase assets corresponding to \emph{will replicate} or \emph{will not replicate} outcomes of a notional replication study of the primary claim of that paper. Agent purchase logic is defined using a sigmoid transformation of a convex semi-algebraic set defined in feature space. Asset prices are determined by a logarithmic scoring rule, and for simplicity, agents specialize in purchasing one of the two asset classes. Time-varying asset prices affect the structure of the semi-algebraic sets leading to time-varying agent purchase rules (see \cite{nakshatri2021design} for further detail including theoretical properties of the market). The price of a \emph{will replicate} asset at market close is taken as proxy for confidence in the primary claim of the paper.

During training, parameters that define agent purchase logic are identified using an evolutionary algorithm. The objective function minimizes root mean square error of the estimated score. Agent performance is evaluated by profit made. Profitable agents are retained, allowed to replicate and then modified using mutation and crossover of parameter values. Agents that do not make a profit are deleted. 

\subsection{Explainability}

The current prototype provides explanations of scores through the record of agents participating and trades made. Confidence in the system's assessment of a paper is based on the extent of agent participation. Agents are initialized in different positions within feature space, so the trading patterns of each agent can be explained in terms of their position and the geometry defining their purchase logic.

\section{Evaluation}

Initial testing of our prototype system was done using a collection of known replication projects and outcomes.  In particular, we use the Reproducibility Project Psychology \cite{open2015estimating}, Social Science Replication Project \cite{camerer2018evaluating}, Experimental Economics Replication Project \cite{camerer2016evaluating}, Many Labs \cite{klein2014investigating} and Many Labs 2 \cite{klein2018many}. Collectively, those projects represent primary findings of 192 total papers in the social and behavioral sciences, each labeled either \emph{Replicable} or \emph{Not Replicable}. 

\noindent \textbf{Experimental settings.} Five-fold cross validation was used. Each fold contained 153 training and 39 test points. Initial conditions were fixed over the five folds -- specifically, we seeded 5 agents per market, each was given 5 units of cash, and the initial price of a \emph{will replicate} asset was set to 0.5. The genetic algorithm trained over 50 generations.

\noindent \textbf{Results on scored papers.} 
Our system provides a confidence score for 68 of 192 (35$\%$) of the papers in our set. On the set of scored papers, accuracy is 0.894, precision is 0.917, recall is 0.903, and \textbf{F1} is \textbf{0.903} (macro averages). A sizeable un-scored subset of data (65$\%$) is the trade-off for high accuracy on the scored subset of the data. A test point is un-scored when the system has determined it has insufficient information to evaluate it.

\noindent \textbf{System non-scoring.} Unlike most other machine learning algorithms, the synthetic market does not provide an evaluation for every input. Like its human-populated counterparts, the market is vulnerable to lack of participation \cite{Arrow:2008,tetlock2008liquidity,rothschild2014extent}. Agents will not participate if they have not seen a sufficiently similar training point (paper). This is more common when the training dataset is small; in experiments with larger datasets, we have observed participation increases. Meaningful ways to increase agent participation, including hybrid settings with human participants, are being explored.

\section{Acknowledgements}
We acknowledge support by DARPA W911NF-19-2- 0272. This work does not necessarily reflect the position or policy of DARPA and no official endorsement should be inferred.

\bibstyle{aaai22}
\bibliography{references}

\end{document}